\documentclass[floatfix,superscriptaddress,showpacs,amssymb,10pt,aps,prd,reprint,longbibliography,showkeys]{revtex4-1}

\usepackage{allpurpose}
\usepackage{graphicx,epsfig,amssymb} 
\usepackage{amsmath,amsfonts, times}
\usepackage{bm} 

\usepackage[usenames]{color}
\usepackage{natbib}
\usepackage{soul}
\usepackage{xeCJK}

\usepackage{float}

\begin{document}

\author{Longfei Wang}
\affiliation{School of Physics and Technology, Wuhan University, Wuhan, 430072, China}

\author{Junji Jia}
\email[Corresponding author:~]{junjijia@whu.edu.cn}
	\affiliation{Department of Astronomy $\&$ MOE Key Laboratory of Artificial Micro- and Nano-structures, School of Physics and Technology, Wuhan University, Wuhan, 430072, China}

\title{The brachistochrone problem for a constant velocity traveler in static and spherically symmetric spacetimes}

\begin{abstract}
This work investigates the brachistochrone problem for a traveler with constant local velocity within static and spherically symmetric (SSS) spacetimes. The brachistochrone trajectory (BT) equations for ultra-relativistic travelers are derived for general SSS metrics, and the solution is formally obtained in an integral form. We then apply the result to two representative spacetimes corresponding to the singular isothermal sphere (SIS) with a finite boundary and the relativistic Plummer mass profile, respectively. For the SIS spacetime, the BT inside the boundary is solved analytically and found always to bend. As the equation of state index $w$ increases, the turning radius $r_0$ of the BT, and consequently the total time, also increase. For the BT outside the boundary, it is found that the heavier the central object, the larger the $r_0$ and the total travel time. For the relativistic Plummer model, the BT will be a straight line passing through the origin when the initial and final points' radii are comparable or smaller than the size of the core region of the mass distribution. When the end points lie well outside the core region, the BT bends, exhibiting a larger turning radius for a more concentrated core. We then extend the consideration to travelers with subluminal constant velocity $v$ and show through the generalized Fermat's principle that the BT will be the same geodesic in the optical metric as ultra-relativistic travelers, with the total travel time scaled by a factor of $1/v$. 
\end{abstract}


\maketitle

\section{Introduction}

The brachistochrone problem (BP) dates back a long time in the history of science, particularly in mathematics and physics. The cycloid solution of the BP for the descending particle by J. Bernoulli laid the foundation of variational calculus  \cite{goldsteinhistory}. The BP for light propagation led to the famous Fermat principle. In Lorentzian spacetime, this principle forces the brachistochrone trajectory (BT) of a light ray connecting two spatial points to be a straight line. 

In more recent years, the study of BP has evolved in several aspects. The first is that people extended the forms and nature of the forces on the moving particle, from pure uniform gravity to more arbitrary forces, including non-uniform forces \cite{elecbp}, velocity dependent forces \cite{veloforce}, and forces with friction  \cite{Denman:1985remarks,lipp1997,vratanar1998,giambo:2000} etc, and from pure gravity to forces of different nature, especially the electromagnetic force \cite{goldstein1986,elecbp,magbp}. The second is that after special relativity was established, people started to investigate the relativistic BP  \cite{goldstein1986,rita2006}. The third is to extend the study to the general relativistic realm, i.e., considering the motion in gravity described by curved spacetimes  \cite{kamath:1988,perlick1991brachistochrone}.  Since the ``time'' coordinate becomes observer-specific in relativity, the BP naturally depends on the reference frame too. In such cases, usually two types of time are of interest: the general time coordinates specific to a remote and often static/stationary observer and the local time specific to the traveler. Studies to minimize the latter are often termed the proper time  \cite{goldstein1986,perlick1991brachistochrone} or travel time  \cite{Giannoni:1999morse, Piccione:2000timeExtremal} BP. In more recent years, the classical BP is also generalized to the quantum setting  \cite{Assis:Fring:2007brachistochrone,qbp1,qbp2,qbp3,qbp4,qbp5}. 

In this work, we consider the BP in arbitrary static and spherically symmetric (SSS) spacetimes. We focus on the case that the traveler always travels at a constant fraction of the speed of light, although this fraction can approach one. Previously, the BP in curved spacetime was first studied by Kamath in the weak field limit  \cite{kamath:1988}. Perlick then studied the traditional BP and the proper time BP in stationary spacetime for a moving object with constant energy under the influence of gravity only  \cite{perlick1991brachistochrone}. In particular, it was found that the BT minimizing the proper time is nothing but the geodesic curve connecting the initial and end points, which is well expected. Piccione and Giannoni et al. then generalized the situation to cases in which the traveler can also be subject to some constraint forces  \cite{Giannoni:1999morse, Piccione:2000timeExtremal, giambo:2000}, in which case the proper time BTs are naturally no longer geodesic. 
In the current work, unlike these previous considerations, which usually fix the conserved energy of the motion, we will not put an explicit constraint on the energy of the traveler. Instead, we fix the local velocity at which the traveler moves. This is more in line with the scenario that somehow the traveler can reach a fixed cruise speed with respect to the medium it moves through. 

The paper is organized as follows. In Sec. \ref{sec:msp}, we derive the brachistochrone equations (BE) in the SSS spacetime and carry out some preliminary analysis of these equations. We will focus on the solutions to the BE for a trajectory connecting two points with the same radial coordinates, especially two pole points. In Sec. \ref{subsec:in}, we will study the trajectory within the singular isothermal sphere (SIS). In Sec. \ref{subsec:out}, the BT outside the sphere boundary in the Schwarzschild vacuum is then solved. In Sec. \ref{app:plummer}, we will solve the BT in a more realistic mass density, the relativistic Plummer mass profile. Sec. \ref{sec:fermat} extends the consideration to travelers with subluminal speed and discusses the relation between the BP here and Fermat's principle. We conclude the work in Sec. \ref{sec:disc}. Throughout the paper, the natural units $G=c=1$ and the signature convention $(-,\,+,\,+,\,+)$ are used.

\section{Brachistochrone Equation
\label{sec:msp}}

We consider the motion of a traveler in an SSS spacetime, whose metric can always be expressed as
\begin{align}
    \dd s^2=-A(r)\dd t^2+B(r)\dd r^2+C(r)(\dd\theta^2+\sin^2\theta\dd{\varphi}^2)
    \label{equ:spacetime}
\end{align}
where $(t,\,r,\,\theta,\,{\varphi})$ are the Boyer-Lindquist coordinates and $A,\,B,\,C$ ,are functions of $r$ only. We will label the BT with the proper time $\lambda$ and denote the initial and final proper times as $\lambda_i$ and $\lambda_f$, respectively. Then the coordinate time of the trajectory becomes
\begin{align}
    \Delta t =  \int_{0}^{\lambda_f} \dot{t}\,\dd \lambda \label{eq:Tdef}
\end{align} 
where the dot represents the derivative with respect to $\lambda$ and, without losing any generality, we have shifted the initial proper time $\lambda_i$ to $0$. In this section, we will assume that the traveler can keep accelerating itself to an ultra-relativistic speed so that its local velocity can infinitely approach the speed of light. The case that the traveler can only reach a lower constant speed will be studied in Sec. \ref{sec:fermat}. Consequently, in this section the traveler's trajectory can be assumed null. Then using $\dd s^2=0$ in Eq. \eqref{equ:spacetime}, we can find that
\begin{align}
    \dot t = \sqrt{\frac{B(r)\dot r^2 + C(r)\bigl(\dot\theta^2+\sin^2\theta \dot{\varphi}^2\bigr)}{A(r)}}.
        \label{eq:timeterm} 
\end{align}
Substituting this into Eq. \eqref{eq:Tdef}, it is clear that we can treat $\Delta t$ as an effective action while the $\dot t$ as its effective Lagrangian
\begin{align}
    L(r,\theta,{\varphi},\dot r,\dot\theta,\dot{\varphi}) \equiv \dot t=\frac{S(r)}{\sqrt{A(r)}}
    \label{eq:ldef}
\end{align}
where, for notation simplicity, we defined
\begin{align} S(r)=\sqrt{B(r)\dot r^2 + C(r)\bigl(\dot\theta^2+\sin^2\theta \dot{\varphi}^2\bigr)},\label{eq:sdef}
\end{align} 
and then carry out the variational procedure of $\Delta t$ with respect to $\lambda$ to determine the BE.
First, we shall define the generalized momentum, 
\begin{subequations}
\label{eq:p_equation}
\begin{align}
p_r       &\equiv \frac{\partial L}{\partial \dot r} = \frac{1}{\sqrt{A(r)}}\frac{B(r)\dot r}{\sqrt{S(r)}}, \label{eq:prdef} \\
p_\theta  &\equiv \frac{\partial L}{\partial \dot\theta} = \frac{1}{\sqrt{A(r)}}\frac{C(r)\dot\theta}{\sqrt{S(r)}} ,\label{eq:pthetadef} \\
p_{\varphi}    &\equiv \frac{\partial L}{\partial \dot{\varphi}} = \frac{1}{\sqrt{A(r)}}\frac{C(r)\sin^2\theta \dot{\varphi}}{\sqrt{S(r)}}  .\label{eq:pphidef}
\end{align}
\end{subequations}
The BE is then nothing but the corresponding Euler-Lagrangian equation
\begin{align}
    \frac{\dd  p_i}{\dd \lambda}  = \frac{\partial L}{\partial x^i} ,\quad (i=r,\theta,{\varphi}).
\end{align} 
After substituting Eqs. \eqref{eq:ldef} and \eqref{eq:p_equation} for $L$ and the momentum respectively, we obtain, for the $r,\,\theta$ and ${\varphi}$ directions, the following BEs
\begin{subequations}
\label{eq:8}
\begin{align}
&\frac{\dd}{\dd\lambda}\biggl(\frac{B(r)\dot r}{\sqrt{A(r)S(r)}}\biggr) +\frac{A'(r)}{2A(r)^{3/2}}\sqrt{S(r)}-\frac{1}{2\sqrt{A(r)S(r)}}  \notag \\
&~~~~~~~~\times\biggl( B'(r)\dot r^2 + C'(r)\bigl(\dot\theta^2+\sin^2\theta \dot{\varphi}^2\bigr) \biggr) = 0,
\label{eq:8a}\\
&    \frac{\dd }{\dd \lambda}\left(\frac{C(r)\dot\theta}{\sqrt{A(r)S(r)}}\right) - \frac{C(r)\sin\theta\cos\theta\dot{\varphi}^2}{\sqrt{A(r)S(r)}} = 0,
    \label{eq:theta equation}\\
&    \frac{\dd }{\dd \lambda}p_{\varphi} = 0
   . \label{eq:phieq}
\end{align}
\end{subequations}
Here and henceforth, we use the prime to denote the derivative with respect to $r$. Since the Lagrangian \eqref{eq:ldef} is a homogeneous function of all three derivatives $(\dot{r},\,\dot{\theta},\,\dot{{\varphi}})$, only two of the three equations in Eq. \eqref{eq:8} are independent, which we will choose to be \eqref{eq:theta equation} and \eqref{eq:phieq}. The main target from now on becomes to solve these two equations to obtain the BT and corresponding total time. 

We start solving from the simplest BE, i.e., Eq. \eqref{eq:phieq}. Since the effective Lagrangian does not depend on ${\varphi}$, this equation allows a conserved angular momentum solution $p_{\varphi}=C_{\varphi}$. On the other hand, without losing any generality, in SSS spacetime we can always set the initial point to be on the $+\hat{z}$ axis so that at this point $\sin\theta=0$ and consequently from Eq. \eqref{eq:pphidef}, $p_{\varphi}=0$. Then Eq. \eqref{eq:phieq} implies that $p_{\varphi}$ has to be zero along the entire trajectory, which after using Eq. \eqref{eq:pphidef} again, indicates that the only possible solution is $\dot{{\varphi}}=0$. This means that the trajectory is within the plane containing the poles. In other words, the motion degenerates to a two-dimensional one in the $(r,\,\theta)$ space and the problem to determine the BT becomes to solve the $r(\theta)$ or $\theta(r)$ dependence.

Substituting $\dot{{\varphi}}=0$ into Eq. \eqref{eq:theta equation}, it can be integrated once to yield
\begin{align}
    \frac{C(r)\dot\theta}{\sqrt{A(r)S(r)}} = C_1
    \label{eq:conserved_C1}
\end{align}
where $C_1$ is a constant. Substituting $S(r)$ in Eq. \eqref{eq:sdef}, this equation can be transformed to
\begin{align}
    \left(\frac{\dd \theta}{\dd r}\right)^2 =  \frac{C_1^2 A(r) B(r)}{C(r)\bigl[C(r) - C_1^2 A(r)\bigr]}.
    \label{eq:dr_dphi_sq}
\end{align}
The constant $C_1$ here can be linked to the radius $r_0$ of the trajectory at the turning point defined by
\begin{align}\left. \frac{\dd r}{\dd \theta} \right|_{r=r_0} = 0.
\end{align}
Substituting this into Eq. \eqref{eq:sdef}, it is found that $S(r_0)=C(r_0)\dot\theta^2(r_0)$ and further into Eq. \eqref{eq:conserved_C1}, one obtains
\begin{align}
C_1^2 = \frac{C(r_0)}{A(r_0)}.
    \label{eq:C1_sq_val}
\end{align}
Substituting the above into Eq. \eqref{eq:dr_dphi_sq}, we obtain a first-order equation of the motion
\begin{align}
    \frac{\dd \theta}{\dd r} = \pm\sqrt{\frac{A(r) B(r) C(r_0)}{C(r)\bigl[C(r) A(r_0) - C(r_0) A(r)\bigr]}}.
    \label{eq:differential equations}
\end{align}
The plus and minus signs here correspond respectively to segments of the trajectory where $r$ increases or decreases with the increase of $\theta$. Using this, the BT finally can be solved as
\begin{align}
    \theta(r) = \theta_0 + \int_{r_0}^{r} \frac{\dd \theta(r')}{\dd r'} \,\dd r' .
    \label{eq:thetarsol}
\end{align}
where we used $\theta_0$ to stand for $\theta(r_0)$.

\section{BT in SIS spacetime}

In this section, we consider the BT for a traveler in a spacetime of the SIS with a finite boundary. It turns out that the BT within the SIS boundary can be solved analytically.

The metric of this spacetime can be solved from the Tolman-Oppenheimer-Volkoff equation of the isothermal fluid  \cite{Dadhich:2015} with the density $\rho(r)$ and equation of state (EOS) of the form  \cite{Remmen:2021}
\begin{align}
\rho(r)= \frac{w}{ \left(1+6w+w^2\right)}\frac{1}{2\pi r^2},\; P=w\rho,\; 0<w\leq 1,
\end{align}
where $w$ is the EOS index. The resultant metric takes the form \cite{Remmen:2021} 
\begin{align}
\label{eq:sis_metric_2d}
\dd s^2 =
\begin{cases}
-\dfrac{1}{\alpha^2} \left( \dfrac{r}{R} \right)^{\frac{4w}{1+w}} \dd t^2 + \alpha^2 \dd r^2 + r^2 \dd \Omega^2\\
\hspace{5.5cm}r \leq R,
\\
-\left(1-\dfrac{2M}{r}\right) \dd t^2 + \left(1-\dfrac{2M}{r}\right)^{-1} \dd r^2 + r^2 \dd \Omega^2\\
\hspace{5.5cm}r > R,
\end{cases}
\end{align}
where, for the sake of notation, we introduced
\begin{align}
    \alpha = \frac{\sqrt{1+6w+w^2}}{1+w}, 
\end{align}
and $R$ is the radius at which the SIS is truncated because of its infinite mass otherwise. Outside the sphere, it is connected to the Schwarzschild solution and therefore $w$ and $M,\, R$ are related by
\begin{align}
M=m(R)=\int_0^R{4 \pi r^2\rho}\dd r=\frac{2wR}{1+6w+w^2}.
\label{SIS_M}
\end{align}
Since this spacetime has a boundary $R$ separating the matter from vacuum, we will concentrate on the following two cases. The first is that the entire BT is within the radius $R$ and the second is when the BT is completely outside $R$, i.e., in the Schwarzschild vacuum. The mixed case in which the BT might enter or exit the sphere will be skipped for simplicity.

\subsection{BT within the sphere\label{subsec:in}}
Substituting the metric \eqref{eq:sis_metric_2d} into Eq. \eqref{eq:differential equations}, we obtain
\begin{align}
    \frac{\dd \theta}{\dd r} = \pm \frac{\alpha}{r} \left[\left(\frac{r}{r_0}\right)^{2\beta} - 1\right]^{-1/2}
    \label{eq:simple_diff_eq} 
\end{align}
where we introduced
\begin{align}
\beta=\frac{1-w}{1+w}\quad (0\leq\beta<1). \label{eq:betadef}
\end{align} 
Eq. \eqref{eq:simple_diff_eq} immediately suggests that for the BT we considered for a near-null traveler in this spacetime, $r_0$ can only be a minimal but not maximal radius. This is a manifestation that in this spacetime there is no bound orbit for null signals.
Substituting Eq. \eqref{eq:simple_diff_eq} into Eq. \eqref{eq:thetarsol}, we find that the integral can be carried out to yield the BT equation
\begin{align}
    \theta(r) - \theta_0 = \pm \frac{\alpha}{2\beta}\cdot 2\operatorname{arcsec}\left[\left(\frac{r}{r_0}\right)^{\beta}\right].
\end{align} 
This equation can also be inverted to obtain $r(\theta)$, i.e., the  radius of the BT as a function of the angle, as
\begin{align}
    r(\theta) = r_0 \cos^{-\frac{1}{\beta}} \left[ \beta \left(\theta - \theta_0\right)/\alpha \right].
    \label{eq:final_trajectory_r0}
\end{align}

The constants $r_0$ and $\theta_0$ in this trajectory have to be fixed by the coordinates of the initial point $(r_i,\,\theta_i)$ and final point $(r_f,\,\theta_f)$. As stated before, we can always assume that the traveler starts from a point on the $+\hat{z}$ axis, and then $\theta_i=0$ and $\theta_f=\Delta\theta$ is the separation of the polar coordinates of the two ends. Using Eq. \eqref{eq:final_trajectory_r0}, the end point conditions then become 
\begin{subequations}
\label{eq:sis_boundary_equations}
\begin{align}
r_i =& r_0 \left[ \cos(\beta\theta_0/\alpha) \right]^{-\frac{1}{\beta}},\\
r_f =& r_0 \left[ \cos\bigl(\beta(\Delta\theta - \theta_0)/\alpha\bigr) \right]^{-\frac{1}{\beta}}.
\end{align}
\end{subequations}
These two equations allow us to solve the constants $\theta_0$ and $r_0$ as
\begin{subequations}
\label{eq:r0theta0sol}
\begin{align}
\theta_{0} =& \frac{\alpha}{\beta} \cos^{-1} \left\{1+\left[\cot \frac{\beta\Delta\theta}{\alpha} -\left(\frac{r_i}{r_f}\right)^{\beta} \csc \frac{\beta\Delta\theta}{\alpha} \right]^2\right\}^{-1/2},
\label{eq:sis_theta0_solution}
\\
r_{0} =& r_i \left[ \cos(\beta\theta_{0}/\alpha) \right]^{\frac{1}{\beta}}.
\label{eq:sis_r0_solution}
\end{align}
\end{subequations}
This completes the solution of the BT: once the  $(r,\theta)$ coordinates of the initial and final points are known, substituting into Eq. \eqref{eq:r0theta0sol} and further into Eq. \eqref{eq:final_trajectory_r0}, the trajectory is completely known.

\begin{figure}[htp!]
    \centering
    \includegraphics[width=0.48\textwidth]{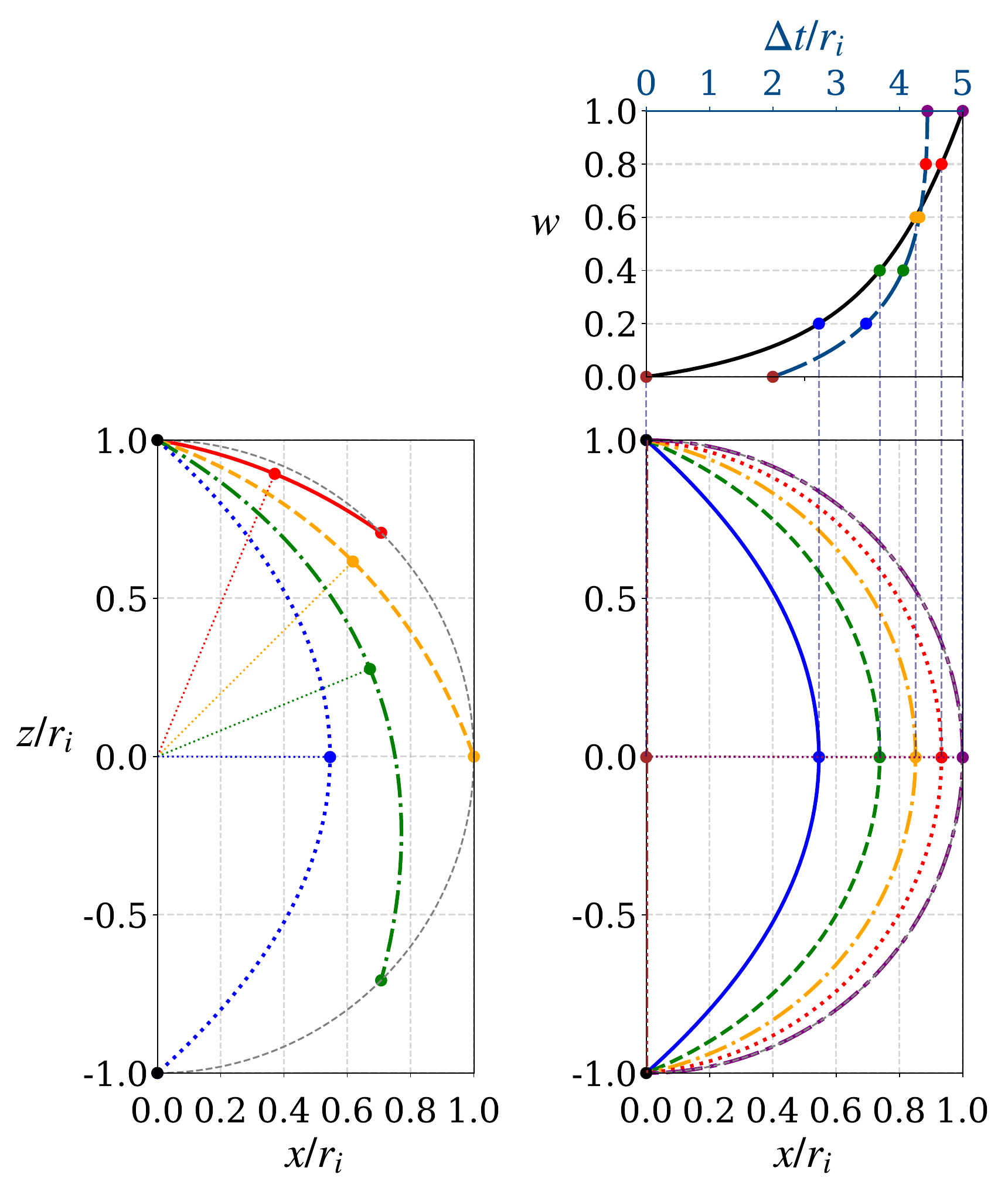}
    \caption{BTs (colored lines) within the SIS connecting: (left panel) end points with same radial coordinates but different angular separation $\Delta\theta=(\pi/4,\,\pi/2,\,3\pi/4,\,\pi)$ for same $w=0.2$, and (right panel) opposite end points for different choices of $w=(0,\,0.2,\,0.4,\,0.6,\,0.8,\,1)$. The top panel on the right side shows the correspondence between $w$ and $r_0$ (solid curve) as well as $w$ and $\Delta t$ (dashed curve). }
\label{fig:sis}
\end{figure}

A few special cases of the BTs deserve to be mentioned here. The first is when the end point has the same radius as the initial point, i.e., $r_f=r_i$. In this case,  according to Eq. \eqref{eq:r0theta0sol},
\begin{align}
    \theta_0=\frac{\Delta\theta}{2}, \quad  r_0 = r_i \biggl[ \cos \Bigl( \frac{\beta \Delta \theta}{2\alpha} \Bigr) \biggr]^{\frac{1}{\beta}}. \label{eq:theta0r0special}
\end{align}
The last equation implies that $r_0$ is always smaller than $r_i$ and therefore the entire BT is within the sphere once $r_i$ and $r_f$ are. Moreover, note that since $0\leq\beta/\alpha<1$ for $0\leq w\leq 1$,  it is also clear that for fixed $w$, $r_0$ monotonically decreases as $\Delta\theta$ increases (see the left plot of Fig. \ref{fig:sis}). On the other hand, if we fix $\Delta\theta$, then it is not too difficult to show that $r_0$ will monotonically increase as $w$ increases.
Substituting Eq. \eqref{eq:theta0r0special} into Eq. \eqref{eq:final_trajectory_r0}, the BT in the $r_i=r_f$ case can be easily obtained as
\begin{align}
    r(\theta) = r_i \left\{ \displaystyle\frac{\cos\left[\displaystyle \frac{\beta}{\alpha} \left(\theta - \frac{\Delta\theta}{2}\right) \right]}{\cos\left( \displaystyle\frac{\beta\Delta\theta}{2\alpha} \right)} \right\}^{-\displaystyle\frac{1}{\beta}}.
\label{eq:rthetaisi}
\end{align}

In Fig. \ref{fig:sis}, we plot the trajectory for this case. In the left panel, we fix $w=0.2$ and plot the BT for several $\Delta\theta$. It is seen that as $\Delta\theta$ increases, the angles for the turning points increase while their radii decrease, both in accord with Eq. \eqref{eq:theta0r0special}. The parameter whose effect on the BT is more interesting is the EOS index $w$. Its effect can be well studied from Eq. \eqref{eq:rthetaisi} after substituting the $\beta$ from Eq. \eqref{eq:betadef}, although the result is tedious. Instead, in the right panel of Fig. \ref{fig:sis}, we fix $\Delta\theta=\pi$, i.e., the end points on the opposite sides, and plot the BT for several $w$. It is clear that as $w$ increases, the $r_0$ increases as predicted after Eq. \eqref{eq:theta0r0special}, and therefore the entire BT for larger $w$ is bent more severely. This actually aligns with our expectation on $w$'s effect. The larger the $w$, the harder the EOS and therefore the denser the matter near the center. In other words, increasing $w$ enlarges the effective refractive index at the same $r$ and therefore the BT has to move around more to achieve the shortest time. In particular, we note from Fig. \ref{fig:sis} that for {\it and only for} vacuum with $w=0$, a straight BT connecting the two poles is obtained.

The coordinate time $\Delta t$ of BT in the SIS spacetime can be obtained from the time integral derived previously from Eq.~\eqref{eq:timeterm}. In particular, after using the trajectory Eq. \eqref{eq:differential equations} and substituting the SIS metric functions, one finds the total coordinate time as
\begin{align}
\Delta t=&2\alpha^2\int_{r_0}^{r_i}
\left(\frac{R}{r}\right)^{\frac{2w}{1+w}}
\frac{\dd r}{\sqrt{1-\left(r_0/r\right)^{2\beta}}} 
\notag \\
=&\frac{2\alpha^2}{\beta}R^{\frac{2w}{1+w}}
\sqrt{r_i^{2\beta} - r_0^{2\beta}} .
\label{eq:SIS_time_integral}
\end{align}
We then used this result to plot $\Delta t$ in Fig. 1 (top right panel) for several $w$. 
It is seen that as $w$ increases, the time $\Delta t$ also increases, which is qualitatively similar to the behavior of $r_0$. However, different from $r_0$, $\Delta t$ starts at $w=0$ from $2r_i$ directly, which is the time to traverse the two pole points along the $z$-axis. As $w$ increases, the rate at which $\Delta t$ increases is slower than that of $r_0$. 

\subsection{BT outside the sphere\label{subsec:out}}

Once the initial and final points are far enough from the isothermal sphere, then the entire BT will be the same as that in the ordinary Schwarzschild spacetime. We briefly study this case in this subsection.

Substituting the Schwarzschild metric into Eq. \eqref{eq:differential equations} and further into Eq. \eqref{eq:thetarsol}, the BT is formally expressed as
\begin{align}
    \theta(r) = \theta_0+\int_{r_0}^{r} \frac{\pm \dd r'}{r'\sqrt{1-\frac{2M}{r'}}\sqrt{\frac{r'^2(1-2M/r_0)}{r_0^2(1-2M/r')}-1}}  . \label{eq:thetarsch}
\end{align}
The $r_0$ in the integral is also the minimal radius along the trajectory. The $\pm$ sign choice in the integral depends on whether the traveler is moving towards or away from the $r_0$ point.
The integral here unfortunately can not be carried out to obtain an elementary function. Consequently, unlike in Subsec. \ref{subsec:in} for BT within the sphere, here we will not be able to invert this relation to obtain the trajectory equation $r(\theta)$.

The $r_0$ and $\theta_0$ in Eq. \eqref{eq:thetarsch} should also be determined by the end point conditions. Since $0<w\leq 1$, from Eq. \eqref{SIS_M} we know that $r>R\geq 4M$. In this region of the radius in Schwarzschild spacetime, it is known that the ultra-relativistic trajectories will not have a bound orbit, and therefore this confirms once again that $r_0$ can only be a minimum but not a maximum.
Again, assuming that the traveler starts from a point on the $+\hat{z}$ axis and $\theta_i=0$ and $\theta_f=\Delta\theta$ is the separation of the polar coordinates of the two ends, the end point conditions then become
\begin{subequations}
\label{eq:sch_boundary_equations}
\begin{align}
0 =& \theta_0+ \int_{r_0}^{r_i} \frac{-1}{r'\sqrt{1-\frac{2M}{r'}}\sqrt{\frac{r'^2(1-2M/r_0)}{r_0^2(1-2M/r')}-1}} \dd r',\\
\Delta\theta =& \theta_0+ \int_{r_0}^{r_f} \frac{\mathrm{sign}(\Delta\theta-\theta_0)}{r'\sqrt{1-\frac{2M}{r'}}\sqrt{\frac{r'^2(1-2M/r_0)}{r_0^2(1-2M/r')}-1}} \dd r'.
\end{align}
\end{subequations}
The $(r_0,\,\theta_0)$ can be numerically solved once $\{\Delta\theta, \, r_i,\,r_f\}$ and and mass $M$ are fixed.

\begin{figure}[htp!]
    \centering

    \includegraphics[width=0.48\textwidth]{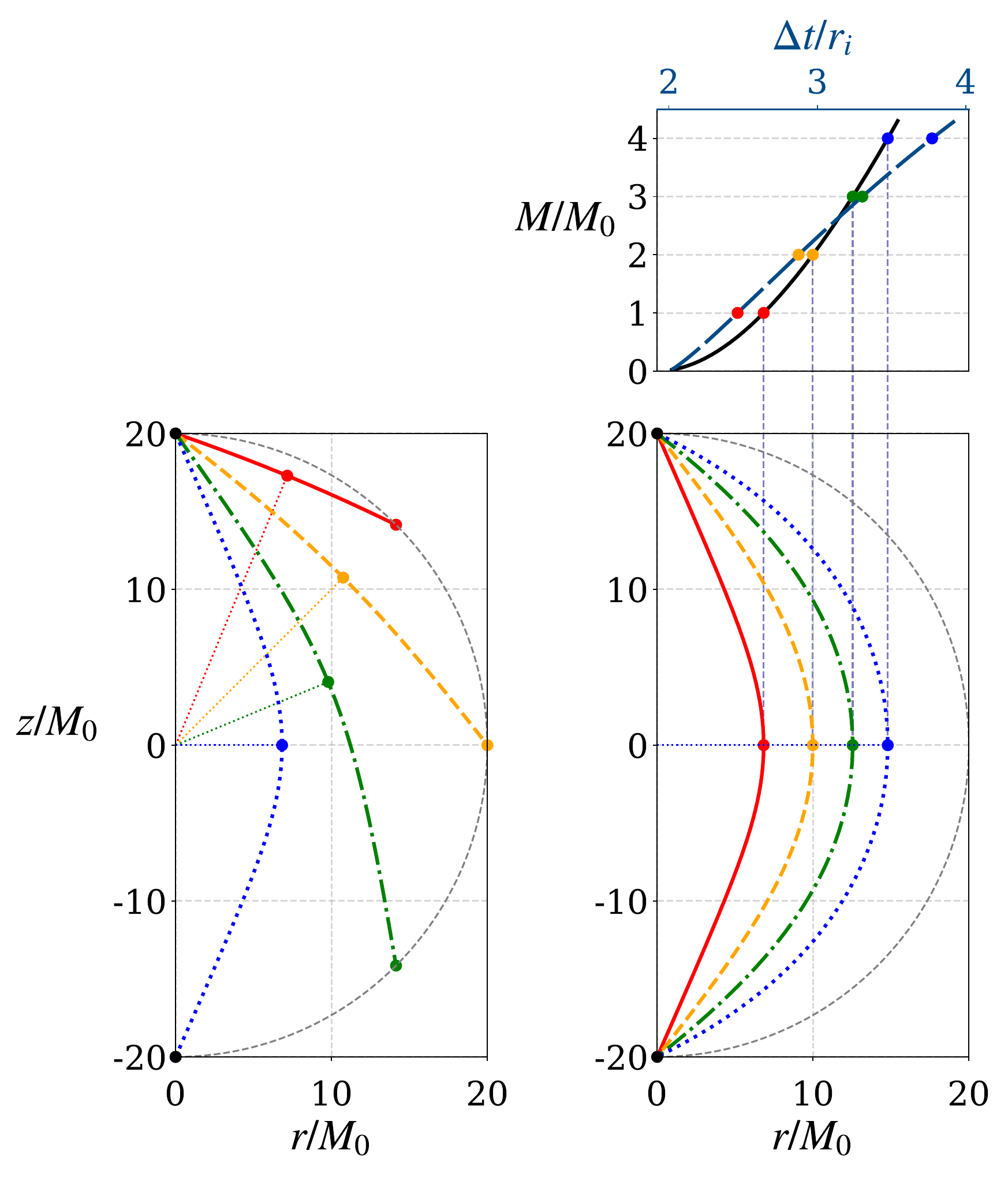}

    \caption{BTs (colored lines) within the Schwarzschild region connecting: (left panel) end points with same radial coordinate but different angular separation $\Delta\theta=(\pi/4,\,\pi/2,\,3\pi/4,\,\pi)$ for same $M=M_0$, and (right panel) opposite end points for different choices of $M=(M_0,\,2M_0,\,3M_0,\,4M_0)$. Here $M_0$ is a base mass/length/time unit we can arbitrarily choose. The top plot on the right panel shows the correspondence between $M$ and $r_0$ (solid curve), as well as $M$ and $\Delta t$ (dashed curve). $r_i=r_f=20M_0$ are used in both plots.}

    \label{fig:sch}

\end{figure}

In Fig. \ref{fig:sch}, we plot the BT for the case of fixed $M$ but different $\Delta\theta$ (left) and the case of fixed $\Delta\theta=\pi$ but varying $M$ (right). It is seen that for the fixed mass $M$ and $r_i=r_f$ case, as $\Delta\theta$ increases, $r_0$ will have to decrease more dramatically than the SIS interior case, as shown in the left panel of Fig. \ref{fig:sis}. This means that to realize the same $\Delta\theta$, the Schwarzschild vacuum case will require a smaller impact parameter, i.e., a closer shooting direction towards the center. This is a reflection of the fundamental physics that the vacuum region of space produces less gravitational attraction. From the right panel of Fig. \ref{fig:sch}, we observe that as $M$ increases, the $r_0$ increases too but with a slower pace. Moreover, the BT also becomes (almost) straight in and only in the zero $M$ limit, which is a similar feature as in the SIS case.

The coordinate time $\Delta t$ of the BT in the Schwarzschild spacetime can be obtained from the time integral derived from Eq.~\eqref{eq:timeterm}. Substituting the Schwarzschild metric and using the trajectory Eq. \eqref{eq:differential equations}, $\Delta t$ can be expressed as
\begin{equation}
\Delta t=2\int_{r_0}^{r_i}
\frac{1}{1-\frac{2M}{r}}
\frac{\dd r}
{\sqrt{
1-\left(\frac{r_0}{r}\right)^2
\left(1-\frac{2M}{r}\right)/\left(1-\frac{2M}{r_0}\right)
}} 
\label{eq:Schwarzschild_time_integral}
\end{equation}
where $r_0$ is determined by Eq.~\eqref{eq:sch_boundary_equations}. Unlike in the SIS case in Eq. \eqref{eq:SIS_time_integral}, the $\Delta t$ here cannot be integrated out and therefore a numerical integration has to be used. In the top-right panel of Fig. \ref{fig:sch}, we plotted the correspondence between $\Delta t$ and different choices of the central mass $M$. It is seen that increasing the mass $M$ results in a longer $\Delta t$. This is indeed expected because, after all, a larger $M$ forces a larger turning radius $r_0$ and a more severely bent trajectory.

\section{BT in the relativistic Plummer galaxy\label{app:plummer}}

In this section, we study the BT in a galaxy whose mass distribution is described by the general relativistic Plummer model. We chose this model because its metric has an explicit analytical form and it is also BH-free and asymptotically flat without having to match the Schwarzschild spacetime outside any boundary. This enables us to study BTs that pass the central region without being trapped by the BH or go to a far region without passing any abrupt boundary. The model is described by the metric  \cite{Tabatabaei:2024}
\begin{equation}
    \dd s^{2} = -\left(\frac{1-\phi}{1+\phi}\right)^{2}\dd t^{2} + (1+\phi)^{4}\left(\dd r^2 + r^2 \dd \Omega^2\right)
    \label{eq:plummer_metric_iso}
\end{equation}
where the potential function
\begin{equation}
    \phi(r) = \frac{1}{2}\frac{M}{\sqrt{r^{2}+b^{2}}}.
    \label{eq:plummer_potential}
\end{equation}
Here $b$ is the Plummer radius, a length scale characterizing the core size, and $M$ is the total mass of the spacetime. In order for the spacetime to be stable everywhere, the constraint $b>M$ has to be imposed, which results in $\phi(r)<1/2$.

To find the BT in this spacetime, substituting the metric \eqref{eq:plummer_metric_iso} into Eq. \eqref{eq:differential equations} and further into Eq. \eqref{eq:thetarsol}, we obtain
\begin{equation}
    \theta(r) = \theta_0+\int_{r_0}^r \frac{\pm\dd r'}{r' \sqrt{\left[\frac{r' \, n(r')}{r_0 \, n(r_0)}\right]^2 - 1}}  \label{eq:thetaplummer}
\end{equation}
where we introduced
\begin{equation}
    n(r) = \frac{(1+\phi(r))^3}{1-\phi(r)}.
    \label{eq:refractive_index_n}
\end{equation}
The integral above can not be carried out to yield a closed form either.

To find the minimal radius $r_0$ as well as the corresponding $\theta_0$, we again use the end points at coordinates $(r=r_i,\theta=0)$ and $(r_f,\theta_f=\Delta\theta)$. Substituting them into Eq. \eqref{eq:thetaplummer} yields
\begin{subequations}
\label{eq:plummer_boundary_equations}
\begin{align}
0 =& \theta_0+ \int_{r_0}^{r_i} \frac{-\dd r'}{r' \sqrt{\left[\frac{r' \, n(r')}{r_0 \, n(r_0)}\right]^2 - 1}}   \dd r',\\
\Delta\theta =& \theta_0+ \int_{r_0}^{r_f}
\frac{\mathrm{sign}(\Delta\theta-\theta_0)\dd r'}{r' \sqrt{\left[\frac{r' \, n(r')}{r_0 \, n(r_0)}\right]^2 - 1}}.
\end{align}
\end{subequations}
Once the boundary parameters $\{r_i,\,r_f,\,\Delta \theta\}$ and the spacetime parameters $\{M,\, b\}$ are known, $(r_0,\theta_0)$ can be solved numerically from these equations. We also point out that all parameters with a length dimension, namely $\{r_0,\,r_i,\,r_f,\,b\}$, can be scaled by a common length variable, which we choose to be the mass $M$, so that the solved $r_0/M$ becomes a function of $r_{i,f}/M$ and $b/M$.

\begin{figure}[htp!]
    \centering
\includegraphics[width=0.48\textwidth]{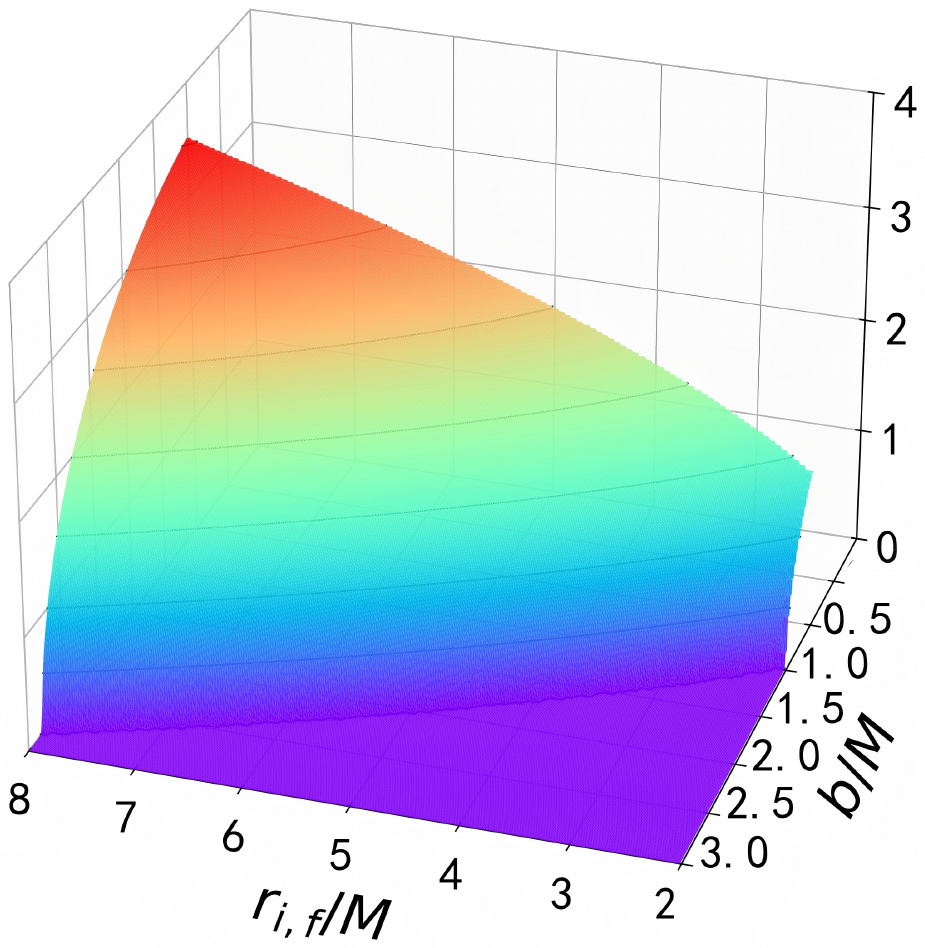}
\caption{Minimal radius $r_0$ for BT connecting the two poles in the relativistic Plummer galaxy, as functions of $b/M$ and $r_{i,f}/M$. }
    \label{fig:plummer 3D}
\end{figure}

To illustrate the results for $(r_0,\,\theta_0)$ and the corresponding BT, we fix the boundaries to be at the opposite poles with the same radial coordinates, i.e.,  $r_i=r_f$ and $\Delta\theta=\pi$.
The corresponding $\theta_0$ is found to be exactly $\pi/2$, as expected from the symmetry of the spacetime and endpoints. For $r_0$, we plot the solved value from Eq. \eqref{eq:plummer_boundary_equations} in  Fig. \ref{fig:plummer 3D} as functions of the parameters $b/M$ and $r_{i,f}/M$. It is seen that for any constant $b/M$, as $r_{i,f}/M$ decreases, the minimal $r_0$ decreases monotonically. Moreover, at a value of $r_{i,f}$ roughly the same order as $b $, the $r_0$ becomes zero identically. This implies that when the initial and final points of the trajectory are within the core radius of the relativistic Plummer mass distribution, the straight line directly connecting the two pole points through the center is the one with the shortest time. If the trajectory starts and ends well outside the core, then a bent trajectory with nonzero $r_0$ shortens the time. This is a key feature that is qualitatively different from that in the SIS mass distribution or Schwarzschild vacuum, where the BT is always bent as seen from Fig. \ref{fig:sis} and Fig. \ref{fig:sch}. For fixed and large enough $r_{i,f}$ but varying $b/M$, we also see that the smaller the $b/M$, the larger the $r_0/M$. This implies that if $M$ is fixed, then a smaller $b$, which corresponds to a more concentrated mass distribution, i.e., a smaller but denser core region, will require a larger $r_0$, resulting in a more bent BT. This feature is qualitatively similar to what is observed in the SIS model in the last section with increasing index $w$ there.

\begin{figure}[htp!]
    \centering
    \includegraphics[width=0.48\textwidth]{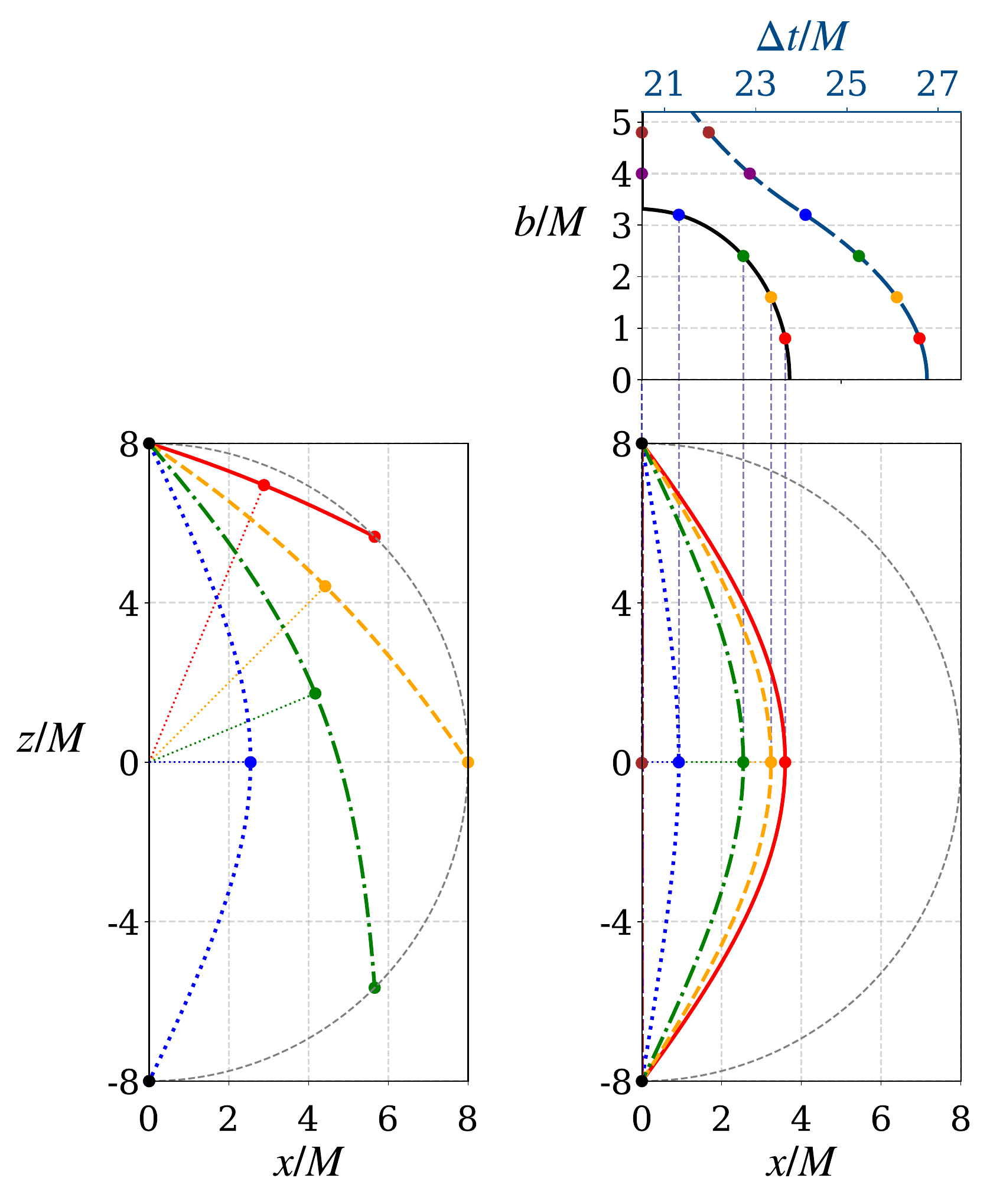}
    \caption{BTs (colored lines) within the relativstic Plummer spacetime connecting: (left panel) end points with the same radial coordinate but different angular separations $\Delta\theta=(\pi/4,\,\pi/2,\,3\pi/4,\,\pi)$ for fixed $b/M=2.4$, and (main right panel) opposite end points ($\Delta\theta=\pi$) for different choices of $b/M=(0.8,\,1.6,\,2.4,\,3.2,\,4.0,\,4.8)$. The top right inset shows the correspondence between $b$ and $r_0$ (solid curve) and $b$ and $\Delta t$ (dashed curve). }
    \label{fig:plummer_combined}
\end{figure}

In the right panel of Fig. \ref{fig:plummer_combined}, we plot the corresponding BTs for several $b/M$ with fixed $r_{i,f}=8M$ and $\Delta\theta=\pi$. It is seen that, as expected from the left panel, $r_0$ decreases as $b/M$ increases, until $b$ becomes comparable to  $r_{i,f}=8M$, beyond which $r_0$ becomes zero and the BT becomes the straight line (the purple line) connecting the poles. To be complete, in the lower left panel of Fig. \ref{fig:plummer_combined}, we plot the BTs for fixed $r_{i,f}=8M$ and $b=2.4M$ but with different $\Delta\theta$. Similar to the cases in the last section, $\theta_0$ keeps increasing while $r_0$ decreases as $\Delta\theta$ increases.

The total coordinate time $\Delta t$ along the BT in the relativistic Plummer spacetime can be obtained by substituting the trajectory Eq. \eqref{eq:differential equations} and then the metric functions \eqref{eq:plummer_metric_iso} into Eq.~\eqref{eq:timeterm}. The result takes the form
\begin{equation}
\Delta t=2\int_{r_0}^{r_i}
n(r)
\frac{1}
{\sqrt{
1-
\left[
\frac{r_0}{r}
\frac{n(r_0)}{n(r)}
\right]^2
}}
\dd r 
\label{eq:Plummer_time_integral}
\end{equation}
where $r_0$ is determined by the end point condition \eqref{eq:plummer_boundary_equations}. The integration here cannot be carried out either, and therefore we will only show $\Delta t$ obtained numerically in the top-right panel of Fig. \ref{fig:plummer_combined}. It is seen that for fixed end-point radius $r_{i,f}$,  the time $\Delta t$ decreases monotonically as the Plummer radius $b/M$ increases, i.e., the mass becomes less concentrated. This is qualitatively similar to the behavior of $r_0$ under the influence of $b/M$. However, different from $r_0$, we notice that even for the case of $b/M=4.0$ or $4.8$ where $r_0$ has reached zero, and the BT becomes straight, the travel time $\Delta t$ keeps decreasing as the mass becomes more dilute.

\section{Fermat's principle and generalization to arbitrary constant velocity} \label{sec:fermat}

There is an important alternative way to understand and derive the general BE in the SSS spacetime. That is, through Fermat's principle in curved spacetimes. This principle states that the smooth curve connecting two spatial points in a conformally stationary spacetime is a (spatially) null geodesic if and only if it minimizes the coordinate time  \cite{Perlick1990_Part1, Perlick:1990df}.
Consequently, one can verify without much difficulty that Eq. \eqref{eq:differential equations} is consistent with the null geodesic equation with the corresponding normalization condition, i.e.,
\begin{align}
&\frac{\dd^2 x^\alpha}{\dd\lambda^2}+\Gamma^\alpha_{\beta\gamma}\frac{\dd x^\beta}{\dd\lambda}\frac{\dd x^\gamma}{\dd\lambda}=0,\\
&g_{\alpha\beta}\frac{\dd x^\alpha}{\dd\lambda}\frac{\dd x^\beta}{\dd\lambda}=0
\end{align}
in the spacetime described by the metric \eqref{equ:spacetime}.

The current consideration can also be generalized to the case that the traveler can only reach a speed of a constant fraction of the light speed with respect to local static observers. Again, assuming that the traveler still moves within a constant ${\varphi}$, then from the point of view of the local static observer, the speed  $v(r,\theta)$ of the traveler can be expressed using the ratio of the spatial element $\dd l$ to the local proper time $\dd\lambda$. Using metric \eqref{equ:spacetime}, we have
\begin{align}
v(r,\theta)^2 =&\frac{\dd l^2}{\dd\lambda^2}= \frac{B(r)\dd r^2 + C(r) \dd\theta^2 }{A(r)\dd t^2}.
\label{eq:local_velocity}
\end{align}
Then the total time is still defined Eq. \eqref{eq:Tdef}, which after using Eq. \eqref{eq:local_velocity} becomes
\begin{align}
    \Delta t =& \int_{t_i}^{t_f} \dd t\nn\\
    =&\frac{1}{v} \int_{\theta_i}^{\theta_f}  \sqrt{\frac{B(r)}{A(r)} \left(\frac{\dd r}{\dd \theta}\right)^2 + \frac{ C(r)}{A(r)}} \, \dd\theta
\label{eq:time_functional_theta}
\end{align}
where a constant $1/v$ was factored out. Comparing to Eq. \eqref{eq:Tdef}, clearly this is only a constant times the travel time $\Delta t$ there. One can then show that the variational procedure would eventually result in the same BT equation as Eq. \eqref{eq:dr_dphi_sq}. Consequently, for the same end points as the boundary condition, these equations lead to the same BT, but the total time is $1/v$ times larger now. We also point out that this coincidence of the spatial part of the BT of a traveler with speed $v$ with the null geodesic of the optical metric can also be viewed as the necessary and sufficient condition of the generalized Fermat's principle in stationary spacetime for massive signals  \cite{alsing}. We note that the $x^i(t)~(i=r,\theta,\varphi)$ functions for these two cases will not be the same though. 

We further remind the reader that although the BT for the traveler with a constant speed $v<c$ can also be thought of as a geodesic, it is not the geodesic in the four-dimensional spacetime of a timelike traveler with some kind of subluminal velocity. Instead, the geodesic curve is still the same null trajectory in the optical metric. This should not be confused with the conclusion that the {\it proper time BT} under pure gravity in stationary spacetime is just the geodesic line  \cite{perlick1991brachistochrone}.

\section{Conclusions} \label{sec:disc}

In this work, we investigate the BT for a traveler with a constant local velocity that is either ultra-relativistic or only a fraction of the speed of light in general SSS spacetimes. We first derived the general BT equation for an ultra-relativistic traveler and showed that the trajectory becomes two-dimensional, which we chose to include the $\hat{z}$ axis. We solved the BT equation in the interior of the SIS spacetime, the exterior Schwarzschild spacetime, and a relativistic Plummer mass profile. It is found that for the SIS and Schwarzschild spacetimes, the BTs always bend, and the larger the EOS index $w$ or mass $M$ in these spacetimes, the more bent the BT becomes. The total coordinate times in these cases also increase as these parameters increase. 

In the relativistic Plummer profile, when the initial and final points are well outside the central region characterized by the Plummer radius $b$, then the BT is also bent. On the contrary, if $b$ becomes comparable to or larger than the end point radii, then the BT becomes a straight line passing through the origin and directly connecting the pole points. The total coordinate time keeps decreasing as $b/M$ increases, even after the BT becomes straight.  

When the traveler speed is only a constant $v$ fraction of that of light, then we showed using the generalized Fermat's principle that the BT would be the same as those for light-speed travelers, although the cost of total time would be $1/v$ times larger. 

The current consideration can be extended in a few interesting ways. The first is to extend the spacetimes to less symmetric ones, such as those that are only axisymmetric. There, the BT is expected to be generally non-planar and therefore more involved. An even more difficult extension is to consider the problem in non-stationary spacetimes. The BEs, which are obtained using a variational principle, would be complicated by the time dependence of the spacetime. We are currently exploring some of these directions.

\section*{Acknowledgements}

L. Wang is supported by the Undergraduate Training Program for Innovation of Wuhan University (No. S202610486124). This work is partially supported by a research development fund from Wuhan University.

\end{document}